\documentclass[aps,prl,twocolumn,superscriptaddress,amsmath,linenumbers, groupedaddress]{revtex4} 
\usepackage[T1]{fontenc}
\usepackage{graphicx} 
\usepackage{dcolumn} 
\usepackage{colordvi}
\usepackage{color}
\usepackage{epstopdf}
\usepackage{pstricks}
\usepackage{amssymb}
\usepackage{url}
\graphicspath{{ps}}
\usepackage{hyperref}
\usepackage{tabularx}
\usepackage{multirow}
\usepackage{upgreek}
\usepackage{hyphenat}
\usepackage[caption=false]{subfig}
\usepackage{amssymb}
\usepackage{amsmath}
\usepackage{commath}
\usepackage{graphicx,bm}
\usepackage{verbatim}
\usepackage{orcidlink}

\input{belle2-symbols}

\begin{document}

\begin{flushright}
Belle II Preprint 2023-006\\
KEK Preprint 2023-2
\end{flushright}

\title{Measurement of {\boldmath $C\!P$} violation in {\boldmath $\Bz\rightarrow \KS\piz$} decays at Belle~II}

  \author{I.~Adachi\,\orcidlink{0000-0003-2287-0173}} 
  \author{K.~Adamczyk\,\orcidlink{0000-0001-6208-0876}} 
  \author{L.~Aggarwal\,\orcidlink{0000-0002-0909-7537}} 
  \author{H.~Ahmed\,\orcidlink{0000-0003-3976-7498}} 
  \author{H.~Aihara\,\orcidlink{0000-0002-1907-5964}} 
  \author{N.~Akopov\,\orcidlink{0000-0002-4425-2096}} 
  \author{A.~Aloisio\,\orcidlink{0000-0002-3883-6693}} 
  \author{N.~Anh~Ky\,\orcidlink{0000-0003-0471-197X}} 
  \author{D.~M.~Asner\,\orcidlink{0000-0002-1586-5790}} 
  \author{H.~Atmacan\,\orcidlink{0000-0003-2435-501X}} 
% ...
\collaboration{The Belle II Collaboration}

\begin{abstract}
    \vspace*{20pt}
    We report a measurement of the $\CP$-violating parameters $\ACP$ and $\SCP$ in $\Bz\to\KS\pi^0$ decays at Belle~II using a sample of $387\times 10^{6}$ $\BBbar$ events recorded in $\ep\en$ collisions at a center-of-mass energy corresponding to the $\Upsilon(4S)$ resonance.
These parameters are determined by fitting the proper decay-time distribution of a sample of 415 signal events.
We obtain $\ACP = -0.04^{+0.14}_{-0.15}\pm 0.05$ and $\SCP = 0.75^{+0.20}_{-0.23}\pm 0.04$, where the first uncertainties are statistical and the second are systematic.
\end{abstract} 

\maketitle

The $B^{0}\to \Kz\pi^{0}$ decay proceeds mainly via the $b\to s \ddbar$ loop amplitude, involving emission and reabsorption of a virtual $W$ boson and a top quark, that carries a weak phase arg$\left(V_{tb}V_{ts}^{*}\right)$.
Throughout this paper, charge-conjugate modes are implicitly included.
Here, $V_{ij}$ denotes Cabibbo--Kobayashi--Maskawa (CKM) matrix elements~\cite{CKMmatrix1,CKMmatrix2}.
The decay is suppressed in the Standard Model (SM) due to the smallness of $|V_{ts}|$.
As non-SM particles can potentially propagate in the loop, studies of this decay provide sensitivity to physics beyond the SM.
Such non-SM physics can manifest itself as an asymmetry in the rates of $\CP$-conjugate decays, i.e., $\CP$ violation~\cite{NewPhysics}.

In the $\Bz\to\Kz\piz$ channel, $\CP$ violation results from either interference between two $\Bz$ decay amplitudes, or interference between a $\Bz$ decay amplitude and that of a $\Bzb$ following $\Bz$--$\Bzb$ mixing.
These two phenomena are quantified by the parameters $\ACP$ and $\SCP$, respectively.
The parameter $\ACP$ is also denoted as $-A$ in the literature.
Neglecting subleading amplitudes with a different weak phase and $\CP$ violation in mixing, we expect $\ACP=0$ and $\SCP=\sin 2\phi_{1}$, where $\phi_{1}\equiv$ arg$\left(-V_{cd}V^{*}_{cb}/V_{td}V^{*}_{tb}\right)$.
The parameter $\sin 2\phi_{1}$ is measured to be $0.70 \pm 0.02$~\cite{HFLAV} in decays mediated by $b\to \ccbar s$ transitions such as $B^{0}\to J/\psi\KS$.
However, the contribution from a color- and CKM-suppressed $b\to \uubar s$ tree amplitude, involving the bottom-to-up-quark transition via a $W$ boson emission, introduces an extra weak phase~\cite{BNpaper,CGRZpaper,Jchaipaper,LSpaper,GLNQpaper};
this shifts the $\SCP$ value from $\sin 2\phi_{1}$.
The resulting difference, $\Delta\SCP\equiv\SCP -\sin 2\phi_{1}$, is estimated in a number of theoretical approaches.
Predictions of $\Delta\SCP$ based on QCD factorization range between $0.01$ and $0.12$~\cite{BNpaper,beneke}, while those based on $SU(3)$ symmetry provide a less stringent lower bound of $-0.06$~\cite{CGRZpaper,GLNQpaper,GRZpaper}.
Similarly, the predicted value of $\ACP$ due to the color-suppressed tree amplitude ranges from $-0.01$ to $0.07$~\cite{BNpaper,CGRZpaper}.
Deviations of $\Delta\SCP$ and $\ACP$ from their expected values would indicate either large subleading amplitudes or non-SM physics~\cite{Robert}.

The parameters $\ACP$ and $\SCP$ are determined from the difference between the decay-time distributions of $\Bz\to\KS\piz$ and $\Bzb\to\KS\piz$ decays.
The $\babar$ and Belle experiments have measured these $\CP$ asymmetries using $467\times 10^6$ and $657\times 10^6$ $\BBbar$ ($B=\Bz$ or $\Bp$) events, respectively~\cite{Babar,Belle}.
The corresponding $\ACP$ ($\SCP$) values are $0.13\pm 0.13$ ($0.55\pm 0.20$) and $-0.14\pm 0.14$ ($0.67\pm 0.32$).

In this Letter, we report the first measurement of $\ACP$ and $\SCP$ in the $\Bz\to\KS\piz$ decay from the Belle~II experiment.
We use a sample of $(387\pm 6)\times 10^{6}$ $\BBbar$ events collected in $\ep\en$ collisions at a center-of-mass (c.m.) energy corresponding to the $\Upsilon(4S)$ resonance.

At $e^{+}e^{-}$ experiments operating near the $\Upsilon(4S)$ resonance, pairs of neutral $B$ mesons are coherently produced in the process $e^{+}e^{-}\to \Upsilon(4S)\to \Bz\Bzb$.
When one of these $B$ mesons decays to a $\CP$ eigenstate $f_{\CP}$ such as $\KS\piz$, and the other to a flavor-specific final state $f_{\rm tag}$, the time-dependent decay rate is given by
\begin{widetext}
\begin{eqnarray}
\mathcal{P}(\Delta t, q)=\frac{{\rm e}^{-|\Delta t|/\tau_{\Bz}}}{4\tau_{\Bz}} \Bigl\{1+q\left[\SCP\sin(\Delta m_{d}\Delta t)-\ACP\cos(\Delta m_{d}\Delta t)\right]\Bigl\},
\label{equation:eqn1}
\end{eqnarray}
\end{widetext}
where $\Delta t=t_{\CP}-t_{\rm tag}$ is the difference in proper times between the two decays, $q$ is the flavor of the tag-side $B$ meson ($+1$ for $\Bz$ and $-1$ for $\Bzb$), $\tau_{\Bz}$ is the $\Bz$ lifetime, and $\Delta m_{d}$ is the $\Bz$--$\Bzb$ mixing frequency.
This study employs a time-dependent $\CP$ analysis method similar to previous measurements~\cite{Belle,Babar}.
The important challenge is determining the location of the $\Bz\to\KS\piz$ decay vertex, which is essential for the $\Delta t$ determination, in the absence of any charged particle originating from the vertex.
The analysis is developed and tested with simulation and validated with a control sample of $\Bz\to J/\psi\KS$ decays before examining the $\Bz\to\KS\piz$ candidates in the data.

The Belle II detector~\cite{belle2tdr,belle2ptp} operates at the SuperKEKB asymmetric-energy ($4\gev$ $\ep$ on $7\gev$ $\en$) collider~\cite{supkek}.
The detector consists of several subdetectors surrounding the interaction region in a cylindrical geometry and is divided into two sections depending on the coverage in polar angle $\theta$.
The two sections are the barrel ($32.2^{\circ}<\theta < 128.7^{\circ}$) and endcap ($12.4^{\circ}<\theta<31.4^{\circ}$ or $130.7^{\circ}<\theta <155.1^{\circ}$).
The subdetectors most relevant for our study are a silicon-based vertex detector (VXD), a gas-based central drift chamber (CDC), and an electromagnetic calorimeter (ECL) made of CsI(Tl) crystals.
The VXD is the innermost component, comprising two layers of pixel sensors surrounded by four layers of double-sided strip sensors~\cite{svd-paper}.
The second pixel layer was incomplete, covering one-sixth of the azimuthal acceptance, for the data analyzed here.
The VXD samples the trajectories of charged particles (``tracks'') near the interaction region to determine the decay positions of their parent particles.
The CDC is the main device for track reconstruction and measurements of particle momenta and charges.
The ECL measures photon energies.

We analyze collision data recorded at the $\Upsilon (4S)$ resonance, corresponding to an integrated luminosity of $362\invfb$.
We use large samples of simulated $\Upsilon(4S)\to\BBbar$ and $e^{+}e^{-}\to\qqbar$ $(q=u,d,s,c)$ events to optimize the event selection and study background distributions.
Simulated $\Bz\to\KS\piz$ events are used to model signal decays and calculate the reconstruction efficiency.
We use \textsc{EvtGen}~\cite{evtgen} to generate $\Upsilon(4S)\to\BBbar$ with the subsequent $B$-meson decays and \textsc{Photos}~\cite{photos} to incorporate final-state radiation from charged particles.
The simulation of $\qqbar$ background relies on the \textsc{Kkmc} generator~\cite{kkmc} interfaced to \textsc{Pythia}~\cite{pythia}.
The detector response for final-state particles is simulated with \textsc{Geant4}~\cite{geant}.
Events are reconstructed using the Belle~II software~\cite{BASF2,BASF2_link}.

Candidate $\KS$ mesons are reconstructed from pairs of oppositely charged tracks, which are assumed to be pions and fit to a common vertex.
The resulting invariant mass is required to lie between $489\mev$ and $507\mev$, corresponding to a $\pm 3\sigma$ range around the known $\KS$ mass~\cite{PDG}, with $\sigma$ being the resolution.
We suppress contamination from prompt $\KS$ candidates and $\Lambda$ decays using two boosted-decision-tree (BDT) classifiers~\cite{bdt}.
These BDTs rely mostly on kinematic information from the $\KS$ and its decay products.

Photons are identified as isolated energy deposits in the ECL that are not matched to any track in the CDC.
We reconstruct $\piz$ candidates from pairs of photons that have energies greater than 35 (153)\,\mev if reconstructed in the barrel (endcap) ECL.
The different energy thresholds are used to suppress beam background, which is higher in the endcap than in the barrel section.
We require the diphoton mass to lie between $116\mev$ and $150\mev$ ($\pm 3\sigma$ range in resolution around the $\piz$ mass~\cite{PDG}).
The absolute cosine of the angle between the higher-energy photon’s direction in the $\piz$ rest frame and the $\piz$ direction in the lab frame must also be less than $0.972$.
These criteria reduce contributions from misreconstructed $\piz$ candidates. 
To improve the momentum resolution, we perform a kinematic fit with the diphoton mass constrained to the known $\piz$ mass~\cite{PDG}.

A neutral $B$-meson candidate is reconstructed by combining a $\KS$ candidate with a $\piz$ candidate.
Two kinematic variables are used to select signal $B$ candidates: the beam-energy-constrained mass $(M_{\rm bc})$ and the energy difference $(\Delta E)$.
These are calculated as
\begin{eqnarray}
M_{\rm bc} &=& \sqrt{E^{2}_{\rm beam} - |\vec{p}_{B}|^{2}},\\ \nonumber
\Delta E &=& E_{B} - E_{\rm beam},
\end{eqnarray}
where $E_{\rm beam}$ is the beam energy, and $\vec{p}_{B}$ and $E_{B}$ are the momentum and energy, respectively, of the $B$ meson.
All quantities are calculated in the c.m.\ frame.
Correctly reconstructed signal candidates peak in $M_{\rm bc}$ at the known $\Bz$ mass~\cite{PDG}, and peak in $\Delta E$ at zero.

For $\Bz\to\KS\piz$, the higher-energy photon from the $\piz$ decay causes a significant correlation between $M_{\rm bc}$ and $\Delta E$ due to leakage of energy deposited in the ECL.
To reduce this correlation, when calculating $\vec{p}_{B}$ in Eq.~(2) we replace the magnitude of the $\piz$ momentum with $\sqrt{(E_{\rm beam} - E_{\KS})^{2}- m_{\piz}^{2}}$, where $E_{\KS}$ is the $\KS$ momentum in the c.m.\ frame.
Simulation shows that the modified $M_{\rm bc}$ ($M^{\prime}_{\rm bc}$) reduces the linear correlation coefficient from $19\%$ to $-1\%$ and has an improved resolution over that of $M_{\rm bc}$.
We retain candidate events satisfying $5.24<M^{\prime}_{\rm bc}<5.29~\gev$ and $|\Delta E|<0.30~\gev$. 

To measure the decay-time difference $\Delta t$, we must determine the positions of the signal and tag-side $B$ decay vertices.
These vertices are obtained using information from the position and spread of the $\ep\en$ interaction region, which is modeled as a three-dimensional Gaussian distribution.
The signal $B$ vertex position is obtained by projecting the $\KS$ flight direction, determined from its decay vertex and momentum, back to the interaction region.
The intersection of the $\KS$ flight projection with the interaction region provides a good estimate of the signal $B$ decay vertex, since both the transverse flight-length of the $\Bz$ meson ($\approx 40\,\mu\rm m$) and the transverse size of the interaction region ($\approx 10\,\mu\rm m$) are small as compared to the $\Bz$ flight length along the boost direction ($\approx 140\,\mu\rm m$).
The tag-side vertex is reconstructed with tracks that are not associated with the $\Bz\to\KS\piz$ candidate.
Such tracks must have a minimum momentum of 50 $\mev$ and at least one hit in each of the PXD, SVD, and CDC subdetectors.
We also apply a similar interaction-region constraint as that used for tracks on the signal side.
We approximate $\Delta t$ to be ${\Delta\ell}/\beta\gamma\gamma^{*}$, where $\Delta\ell$ is the distance between signal and tag-side vertices along the $\en$ beam direction, $\beta\gamma$ ($\approx 0.28$) is the Lorentz boost of the $\FourS$ in the lab frame, and $\gamma^{*}$ ($\approx 1.002$) is the Lorentz factor of the $\Bz$ meson in the c.m.\ frame.

We employ a BDT classifier that uses 32 event-topology variables to distinguish the $\qqbar$ background from $B$-meson decays.
The following variables provide the most discrimination: modified Fox--Wolfram moments~\cite{ksfw}, CLEO cones~\cite{cleo}, the thrust value~\cite{thrust} of the rest of the event, and the cosine of the angle between the thrust axis of the signal $B$ and that of the rest of the event.
The BDT is trained on samples of simulated $e^{+}e^{-}\to \qqbar$ and signal events, each equivalent to about three times the size of the dataset.
The BDT outputs a single variable ($C_{\rm BDT}$) that ranges from zero for background-like events to one for signal-like events.
We require $C_{\rm BDT}$ to be greater than $0.6$, which rejects about $93\%$ of the $\qqbar$ background while preserving $80\%$ of the signal.
The remainder of the $C_{\rm BDT}$ distribution strongly peaks near $1.0$ for signal, leading to difficulty in modeling it with an analytic function.
We thus transform it into a new variable, $C_{\rm BDT}^{\prime}={\rm ln}[(C_{\rm BDT}-0.6)/(1.0- C_{\rm BDT})]$, where $0.6$ ($1.0$) is the minimum (maximum) possible value of the remaining $C_{\rm BDT}$ distribution.
The $C_{\rm BDT}^{\prime}$ distribution can be parametrized with a sum of Gaussian functions, and $C_{\rm BDT}^{\prime}$ is later used as a fit variable.

After applying all selection criteria, $3\%$ of the events have more than one $B$ candidate.
Such multiple candidates come from random combinations of final-state particles.
In events with multiple candidates, we choose that with the largest $p$-value resulting from the $\piz$-mass-constrained fit; if that criterion is ambiguous, we select the candidate with the largest $p$-value from the $\KS$-vertex fit.
This selection retains the correct $B$ candidate in $87\%$ of simulated events that have multiple candidates.
The signal efficiency after all selection criteria are applied ($\varepsilon_{\rm rec}$) is $20\%$.
Simulation studies show that $1.7\%$ of signal candidates are incorrectly reconstructed by including a final-state particle from the tag-side $B$ meson.
We consider this small component, arising mostly due to misreconstructed $\piz$, as part of the signal.

The flavor of the tag-side $\Bz$ meson, $q$, is determined from the properties of final-state particles that are not associated with the reconstructed $\Bz\to\KS\piz$ decay.
We use a category-based multivariate flavor-tagging algorithm for this purpose~\cite{flavortagger}.
The algorithm outputs two parameters, the $b$-flavor charge $q$ and $r$, which is an event-by-event tagging quality factor ranging from zero for no flavor discrimination to one for unambiguous flavor assignment.
The dataset is divided into seven $r$ bins that contain similar numbers of events, but have different signal-to-background ratios.

We select events in which $\Delta t$ is well-measured by requiring $|\Delta t|<10.0\,\rm ps$ and $\sigma^{}_{\Delta t}<2.5\,\rm ps$, where $\sigma^{}_{\Delta t}$ is the uncertainty on $\Delta t$, estimated event-by-event.
The $\Delta t$ distribution of these events is fitted to determine $\ACP$ and $\SCP$.
For the remaining events, about $40\%$, the $\Delta t$ distribution is not included in the fit.
However, these events are still useful to constrain $\ACP$, which is sensitive to the relative yields of $\Bz$ and $\Bzb$ decays.
We thus perform a simultaneous extended maximum-likelihood fit to both subsamples in seven $r$ bins~\cite{Babar}.
For each subsample, the likelihood function includes one-dimensional probability density functions (PDFs) for $M^{\prime}_{\rm bc}$, $\Delta E$, and $C_{\rm BDT}^{\prime}$;
for the first subsample, the likelihood also includes a PDF for $\Delta t$ that depends on the flavor tag $q$.
The PDFs for $M^{\prime}_{\rm bc}$, $\Delta E$, and $C_{\rm BDT}^{\prime}$ are taken to be the same for both subsamples, as found in simulation.

The PDFs for the signal component are as follows: $M^{\prime}_{\rm bc}$ is modeled with the sum of a Crystal Ball function~\cite{CB} and a Gaussian function with a common mean; $\Delta E $ with the sum of a Crystal Ball and two Gaussian functions, all three with a common mean; and $C_{\rm BDT}^{\prime}$ with the sum of asymmetric and symmetric Gaussian functions.
The $\Delta t$ PDF is given by
\begin{widetext}
\begin{eqnarray}
\label{equation:tdcpv}
&&\mathcal{P}_{\rm sig}(\Delta t, q)=\frac{{\rm e}^{-|\Delta t|/\tau_{\Bz}}}{4\tau_{\Bz}}\Bigl\{\left[1-q \Delta w_{r}+q\Delta\varepsilon^{}_{{\rm tag},r}(1-2w_{r})\right]+\left[q(1-2w_{r})+\Delta\varepsilon^{}_{{\rm tag},r}(1-q \Delta w_{r})\right] \bigr[\SCP \sin(\Delta m_{d} \Delta t) -\nonumber\\
&& \hspace*{3.7cm} \ACP \cos(\Delta m_{d} \Delta t) \bigr] \Bigl\} \otimes \mathcal{R}_{\rm sig},
\end{eqnarray}
\end{widetext}
where $w_{r}$ is the fraction of wrongly tagged events; $\Delta w_{r}$ is the difference in $w_{r}$ between $\Bz$ and $\Bzb$; $\Delta\varepsilon^{}_{{\rm tag},r}$ is the asymmetry in their tagging efficiencies, which are the fractions of $\Bz$ or $\Bzb$ signal candidates to which a flavor tag is assigned; and $\mathcal{R}_{\rm sig}$ is the $\Delta t$ resolution function.
The resolution function is described by a double Gaussian convolved with an exponential function; the Gaussian means and widths are scaled by $\sigma^{}_{\Delta t}$.
The $\Delta t$ resolution is dominated by the signal-side $\KS$.
Simulation shows that the $\sigma^{}_{\Delta t}$ distributions for signal and background are the same.
We fix $\tau_{\Bz}$ and $\Delta m_{d}$ to the world averages of $1.519\pm 0.004$\,ps and $0.5065 \pm 0.0019$\,$\rm{ps}^{-1}$, respectively~\cite{HFLAV}.
The tagging parameters ($w_{r}$, $\Delta w_{r}$, and $\Delta\varepsilon^{}_{{\rm tag},r}$) are fixed to values obtained from $\Bz\to D^{(*)-}\pi^{+}$ decays~\cite{flavortagger}.
The effective tagging efficiency $\varepsilon_{\rm eff}=\sum_{r}\varepsilon_{{\rm tag},r}(1- 2 w_{r})^{2}$ is $(30.0\pm 1.2)\%$, where $\varepsilon_{{\rm tag},r}$ is the tagging efficiency for the $r$-th bin.
The $w_{r}$ and $\Delta\varepsilon^{}_{{\rm tag},r}$ values are in the ranges $2\%$--$48\%$ and $0.8\%$--$3.6\%$, respectively.
All signal shape parameters are fixed to values obtained from simulation and calibrated with control samples as described below.

For the $\qqbar$ background, an ARGUS function~\cite{AG} is used for $M^{\prime}_{\rm bc}$, a straight line for $\Delta E$, and the sum of asymmetric and symmetric Gaussian functions for $C_{\rm BDT}^{\prime}$.
The $\Delta t$ distribution is modeled with the signal resolution function $\mathcal{R}_{\rm sig}$, as this background is dominated by prompt $\KS$ decays.
We float the $\qqbar$ background yield, ARGUS curvature parameter, and $\Delta E$ slope, but fix the ARGUS endpoint, $C_{\rm BDT}^{\prime}$ and $\Delta t$ shape parameters to the values obtained from the data sideband $5.24<M^{\prime}_{\rm bc}<5.27~\gev$.
All $\qqbar$ shape parameters are taken to be identical for all $r$ bins.

For the $\BBbar$ background, a two-dimensional kernel density estimation PDF~\cite{2D} is used to model the ($M^{\prime}_{\rm bc}$, $\Delta E$) distribution, and the sum of asymmetric and symmetric Gaussian functions is used for $C_{\rm BDT}^{\prime}$. 
The $\Delta t$ distribution is modeled with an exponential function convolved with $\mathcal{R}_{\rm sig}$.
We float the yield of $\BBbar$ background and fix its shape parameters from a fit to the simulated sample.

We correct the common mean and core width of the signal $M^{\prime}_{\rm bc}$, $\Delta E$, and $C_{\rm BDT}^{\prime}$ PDF shapes for possible differences between data and simulation according to values obtained from a control sample of $B^{+} \to \Dzb (\to \KS \piz)\pi^{+}$ decays.
To select these events, we apply the same $\KS$ and $\piz$ criteria as used for the signal channel.
To ensure the similar $\piz$ momentum range for signal and control channels, we require a minimum $\piz$ momentum of $1.5\gev$.
We perform an unbinned maximum-likelihood fit to the distributions of $M^{\prime}_{\rm bc}$, $\Delta E$, and $C^{\prime}_{\rm BDT}$, using PDF shapes similar to those employed to describe the signal decay.

To validate the fitting procedure, we use a control sample of $\Bz\to\jpsi(\to\mu^{+}\mu^{-})\KS$ decays. 
To mimic the signal decay, we do not use information from the two muon tracks to reconstruct the signal $B$ decay-vertex.
We perform an unbinned maximum-likelihood fit to the distributions of $\mbc$ and $\Delta t$, using PDF shapes and resolution functions similar to those employed in the fit to the signal sample. 
The measured $\Bz$ lifetime, $\ACP$, and $\SCP$ are $1.46\pm 0.05$\,ps, $0.10\pm 0.07$, and $0.76 \pm 0.12$, respectively, where the uncertainties are statistical only.
These results are consistent with their world-average values~\cite{HFLAV}, thus validating our $\Bz\to\KS\piz$ fitting procedure.
The above sample is also used to correct the common mean and core width of the resolution function for possible differences between data and simulation.

\begin{figure*}[htb!]
     \subfloat[]{
\includegraphics[scale=0.4]{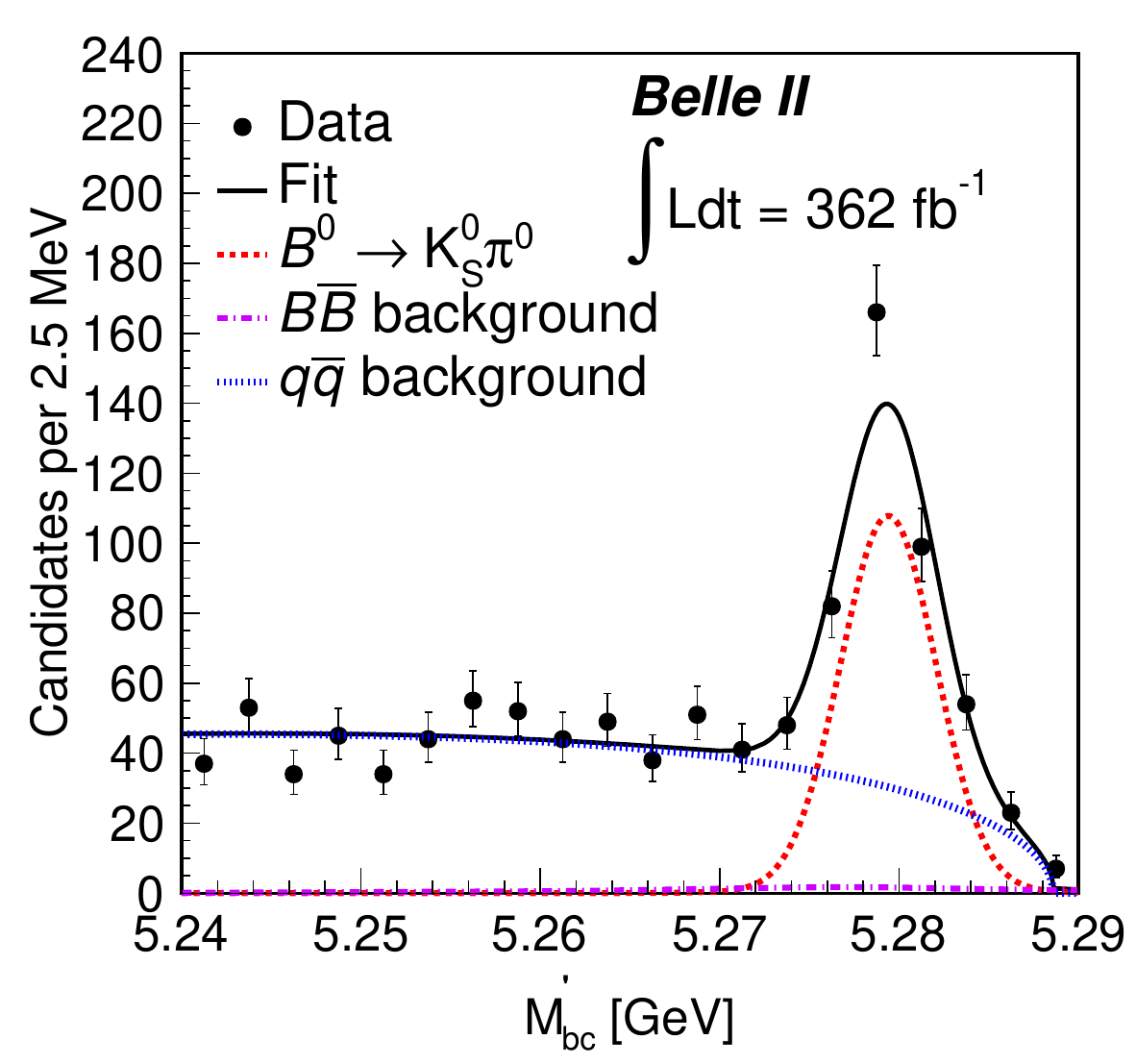}}
     \hfill
      \subfloat[]{
\includegraphics[scale=0.4]{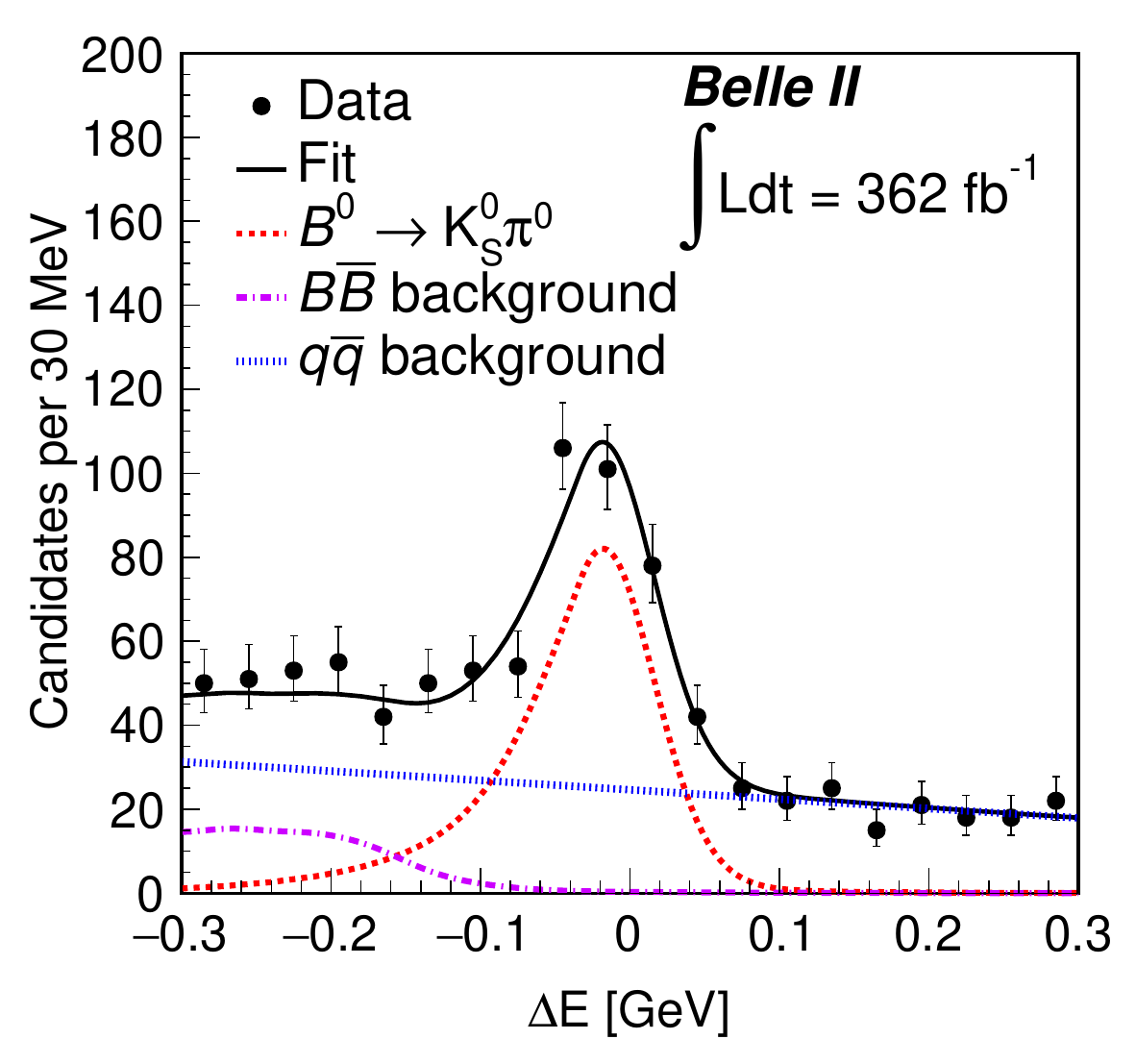}}
     \hfill
      \subfloat[]{
\includegraphics[scale=0.4]{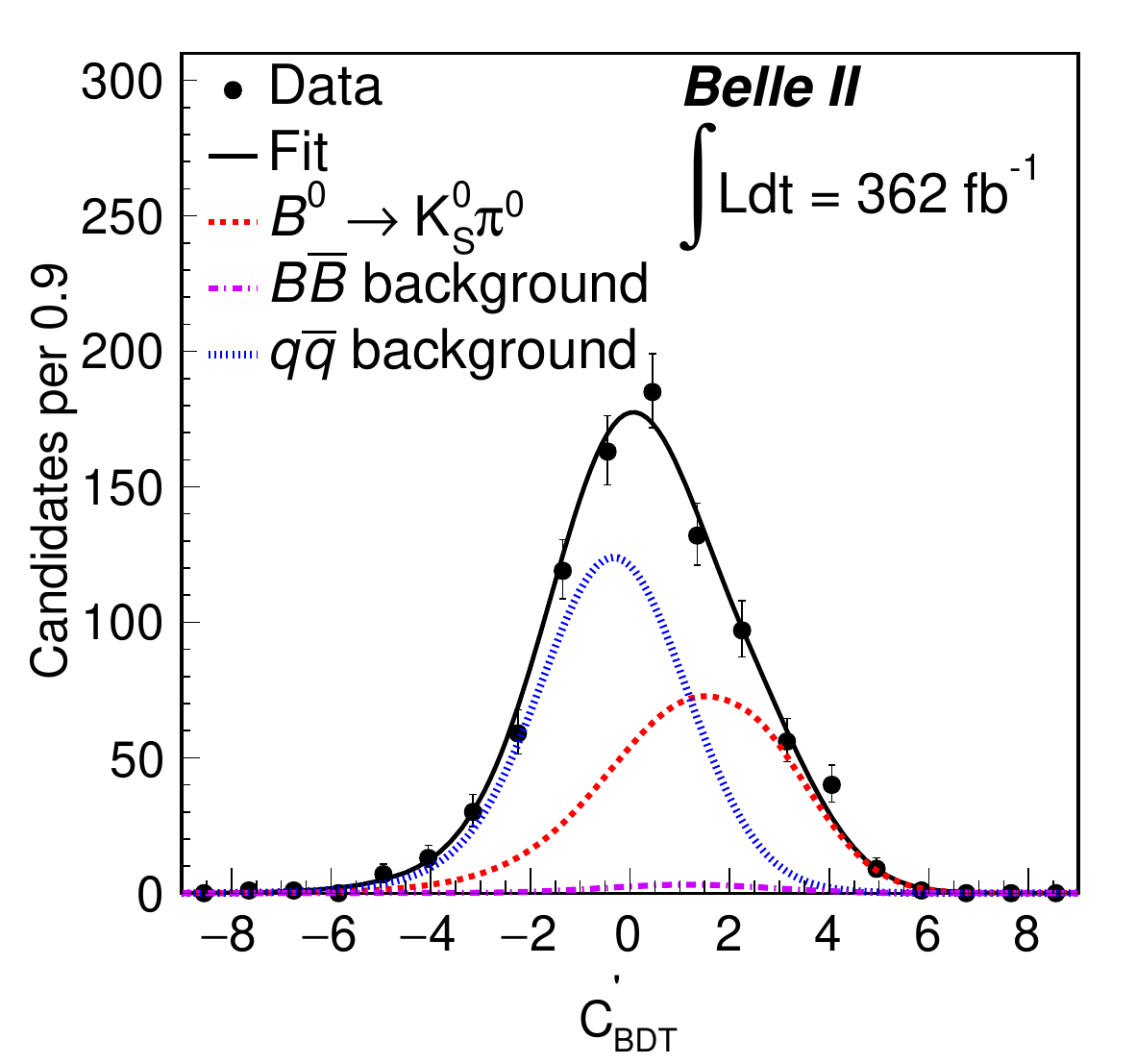}}
       \hfill
        \subfloat[]{
\includegraphics[scale=0.4]{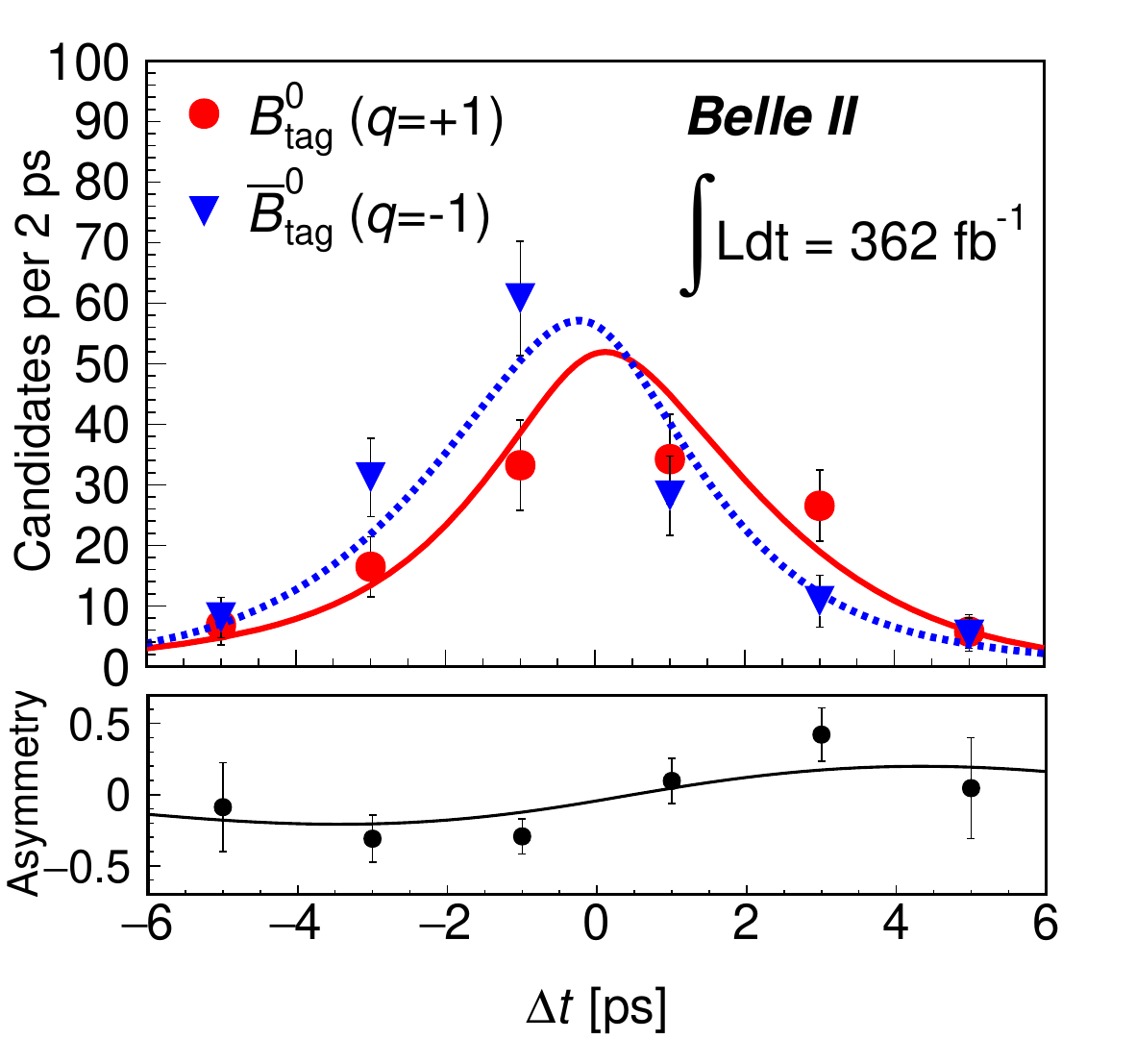}}
     	\caption{Distributions of (a) $M^{\prime}_{\rm bc}$, (b) $\Delta E$, and (c) $C^{\prime}_{\rm BDT}$ with fit projections overlaid for both $\Bz$  and $\Bzb$ candidates satisfying the criteria $5.27<M^{\prime}_{\rm bc}< 5.29\gev$, $-0.15<\Delta E<0.10\gev$, $|\Delta t|<$ 10.0\,ps, and $C^{\prime}_{\rm BDT}>0.0$ (except for the variable displayed). The solid curve shows the fit projection, while various fit components are explained in the legends.
        Distribution of (d) $\Delta t$ for tagged $\Bz$ and $\Bzb$ candidates after subtracting background with the $_s{\mathcal P}$lot method~\cite{sPlot}. The asymmetry, defined as $[N(\Bz_{\rm tag})-N(\Bzb_{\rm tag})]/[N(\Bz_{\rm tag})+N(\Bzb_{\rm tag})]$, is displayed underneath along with the fit projection.}
        
	\label{fig:4D}  
\end{figure*}

Figure~\ref{fig:4D} shows the $M^{\prime}_{\rm bc}$, $\Delta E$, $C^{\prime}_{\rm BDT}$, and $\Delta t$ distributions in the data along with the fit projections overlaid.
For these plots, the seven $r$ bins have been combined, and for all plots except $\Delta t$, both data subsamples (described earlier) are included.
In addition, for each plot the signal-enhancing criteria $5.27<M^{\prime}_{\rm bc}< 5.29\gev$, $-0.15<\Delta E<0.10\gev$, $|\Delta t|<$ 10.0\,ps, and $C^{\prime}_{\rm BDT}>0.0$ have been applied except for the variable displayed.
Distributions of $\Delta t$ with fit projections overlaid are shown in the Supplementary Material~\cite{SupMat}.
The resulting signal yield $N_{\rm sig}$, $\ACP$, and $\SCP$ are $415^{+26}_{-25}$, $-0.04^{+0.14}_{-0.15}$, and $0.75^{+0.20}_{-0.23}$, respectively.
The correlation coefficient between two asymmetries is $-1.7\%$.
From the signal yield, we determine the branching fraction as $\mathcal{B}(\Bz\to\KS\piz)=N_{\rm sig}/(2N_{\BBbar}f^{+0}\varepsilon_{\rm rec})=(11.15^{+0.69}_{-0.67})\times 10^{-6}$, which is consistent with the world average~\cite{HFLAV}.
Here, $f^{+0}$ is the fraction of $\Bz\Bzb$ or $B^{+}B^{-}$ production at the $\Upsilon(4S)$ resonance~\cite{f+0belle} and all quoted uncertainties are statistical.

The systematic uncertainties contributing to $\ACP$ and $\SCP$ are listed in Table~\ref{tab:systematics}.
We estimate the systematic uncertainty due to flavor tagging by individually varying the ($w_{r}, \Delta w_{r}, \Delta\varepsilon^{}_{{\rm tag},r}$) parameters by their uncertainties for each $r$ bin, while considering correlations.
The maximum deviations with respect to the nominal results are taken as systematic uncertainties.
The uncertainty due to the $\Delta t$ resolution function is estimated in a similar fashion.
In the nominal fit, we assume the $\BBbar$ background to be $\CP$ symmetric.
To account for a potential $\CP$ asymmetry in the $\BBbar$ background, we perform a series of fits with the $\Delta t$ PDF formed by varying the $\ACP$ and $\SCP$ values for that background from $-1$ to $+1$ while fixing the effective lifetime value to that determined from simulation.
We then calculate the deviations in signal $\ACP$ and $\SCP$ from their nominal values; the largest deviation is assigned as the systematic uncertainty.
To evaluate the uncertainty due to a possible asymmetry in the $\qqbar$ background, we perform an alternative fit by fixing the asymmetry to that obtained from the data sideband defined earlier.
The uncertainty due to the signal PDF shape is estimated using an alternative model based on kernel-density estimation.
Similarly, the uncertainty due to the background PDF shape is calculated by varying all fixed parameters by their uncertainties and taking the maximum deviation from nominal results as the uncertainty.

A potential fit bias is checked for by performing an ensemble test comprising $1000$ simulated experiments in which signal and $\BBbar$ background events are drawn from simulated samples and $\qqbar$ background events are generated according to their PDF shapes.
We calculate the mean shifts of the fitted values of $\ACP$ and $\SCP$ from their input values and assign them as systematic uncertainties.
The systematic uncertainty due to multiple candidate selection is evaluated by performing an alternative fit with all candidates and taking the difference with respect to the nominal value.
The impact of misreconstructed signal candidates on $\ACP$ and $\SCP$ is negligible.
Uncertainties due to fixed $\tau_{\Bz}$ and $\Delta m_{d}$ values are calculated by varying these quantities by their uncertainties and repeating the fit; the resulting maximum variations in $\ACP$ and $\SCP$ are assigned as systematic uncertainties.
Tag-side interference can arise due to the presence of both CKM-favored and CKM-suppressed tree amplitudes contributing to the tag-side decay~\cite{TSI}.
The resulting impact is conservatively estimated by positing that all events are tagged with such hadronic decays.
The uncertainty due to VXD misalignment is evaluated by reconstructing events with various misalignment hypotheses as done in Ref.~\cite{VXD}.
Assuming all systematic sources to be independent, we add their contributions in quadrature to obtain the total systematic uncertainty of $\pm 0.047$ for $\ACP$ and $\pm 0.040$ for $\SCP$.

\begin{table*}[htb!]
\caption{Systematic uncertainties (absolute) contributing to the time-dependent $\CP$ asymmetries.}
\label{tab:systematics}
\centering
    \begin{tabular}{lccc}
    \hline \hline
  Source & ~~~~~~~$\delta\ACP$  ~~~~~~~& $\delta\SCP$ \\
 \hline
  Flavor tagging & 0.013& 0.011  \\
  $\Delta t$ resolution function &  0.014 & 0.022 \\
  $\BBbar$ background asymmetry  & 0.030 &0.018 \\
  $\qqbar$ background asymmetry  & 0.028 & < 0.001\\
  Signal modeling  & 0.004&0.003 \\
  Background modeling & 0.006&0.018  \\
  Fit bias & 0.005&0.011\\
  Multiple candidate selection &0.005 &0.010 \\
  $\tau_{\Bz}$ and $\Delta m_{d}$ & < 0.001& < 0.001   \\
  Tag-side interference  & 0.006&0.011\\
  VXD misalignment  &0.004 &0.005\\
  \hline 
  Total  &0.047 &0.040\\ \hline\hline
    \end{tabular}
\end{table*}

In summary, we measure the $\CP$-violating parameters $\ACP$ and $\SCP$ in $\Bz\to\KS\piz$ decays using a sample of $387\times 10^{6}$ $\BBbar$ events recorded by Belle II in $\ep\en$ collisions at the $\Upsilon(4S)$ resonance.
Based on a signal yield of $415_{-25}^{+26}$ events, we obtain
\begin{eqnarray}
\ACP=&-0.04^{+0.14}_{-0.15} \pm 0.05
\end{eqnarray}
and
\begin{eqnarray}
\SCP = &0.75^{+0.20}_{-0.23}\pm 0.04,
\end{eqnarray}
where the first uncertainties are statistical and the second are systematic.
This constitutes the first Belle~II measurement of $\CP$ asymmetries in the decay. 
Our results agree with previous determinations~\cite{Belle,Babar}, and the precision obtained for $\SCP$ is better than (similar to) that achieved at Belle ($\babar$), despite using a data sample only $60$--$80\%$ the size of the samples used in those experiments.
The results are consistent with SM predictions and can provide useful constraints on non-SM physics.

This work, based on data collected using the Belle II detector, which was built and commissioned prior to March 2019, was supported by
Science Committee of the Republic of Armenia Grant No.~20TTCG-1C010;
Australian Research Council and research Grants
No.~DP200101792, 
No.~DP210101900, 
No.~DP210102831, 
No.~DE220100462, 
No.~LE210100098, 
and
No.~LE230100085; 
Austrian Federal Ministry of Education, Science and Research,
Austrian Science Fund
No.~P~31361-N36
and
No.~J4625-N,
and
Horizon 2020 ERC Starting Grant No.~947006 ``InterLeptons'';
Natural Sciences and Engineering Research Council of Canada, Compute Canada and CANARIE;
National Key R\&D Program of China under Contract No.~2022YFA1601903,
National Natural Science Foundation of China and research Grants
No.~11575017,
No.~11761141009,
No.~11705209,
No.~11975076,
No.~12135005,
No.~12150004,
No.~12161141008,
and
No.~12175041,
and Shandong Provincial Natural Science Foundation Project~ZR2022JQ02;
the Ministry of Education, Youth, and Sports of the Czech Republic under Contract No.~LTT17020 and
Charles University Grant No.~SVV 260448 and
the Czech Science Foundation Grant No.~22-18469S;
European Research Council, Seventh Framework PIEF-GA-2013-622527,
Horizon 2020 ERC-Advanced Grants No.~267104 and No.~884719,
Horizon 2020 ERC-Consolidator Grant No.~819127,
Horizon 2020 Marie Sklodowska-Curie Grant Agreement No.~700525 "NIOBE"
and
No.~101026516,
and
Horizon 2020 Marie Sklodowska-Curie RISE project JENNIFER2 Grant Agreement No.~822070 (European grants);
L'Institut National de Physique Nucl\'{e}aire et de Physique des Particules (IN2P3) du CNRS (France);
BMBF, DFG, HGF, MPG, and AvH Foundation (Germany);
Department of Atomic Energy under Project Identification No.~RTI 4002 and Department of Science and Technology (India);
Israel Science Foundation Grant No.~2476/17,
U.S.-Israel Binational Science Foundation Grant No.~2016113, and
Israel Ministry of Science Grant No.~3-16543;
Istituto Nazionale di Fisica Nucleare and the research grants BELLE2;
Japan Society for the Promotion of Science, Grant-in-Aid for Scientific Research Grants
No.~16H03968,
No.~16H03993,
No.~16H06492,
No.~16K05323,
No.~17H01133,
No.~17H05405,
No.~18K03621,
No.~18H03710,
No.~18H05226,
No.~19H00682, 
No.~22H00144,
No.~26220706,
and
No.~26400255,
the National Institute of Informatics, and Science Information NETwork 5 (SINET5), 
and
the Ministry of Education, Culture, Sports, Science, and Technology (MEXT) of Japan;  
National Research Foundation (NRF) of Korea Grants
No.~2016R1\-D1A1B\-02012900,
No.~2018R1\-A2B\-3003643,
No.~2018R1\-A6A1A\-06024970,
No.~2018R1\-D1A1B\-07047294,
No.~2019R1\-I1A3A\-01058933,
No.~2022R1\-A2C\-1003993,
and
No.~RS-2022-00197659,
Radiation Science Research Institute,
Foreign Large-size Research Facility Application Supporting project,
the Global Science Experimental Data Hub Center of the Korea Institute of Science and Technology Information
and
KREONET/GLORIAD;
Universiti Malaya RU grant, Akademi Sains Malaysia, and Ministry of Education Malaysia;
Frontiers of Science Program Contracts
No.~FOINS-296,
No.~CB-221329,
No.~CB-236394,
No.~CB-254409,
and
No.~CB-180023, and No.~SEP-CINVESTAV research Grant No.~237 (Mexico);
the Polish Ministry of Science and Higher Education and the National Science Center;
the Ministry of Science and Higher Education of the Russian Federation,
Agreement No.~14.W03.31.0026, and
the HSE University Basic Research Program, Moscow;
University of Tabuk research Grants
No.~S-0256-1438 and No.~S-0280-1439 (Saudi Arabia);
Slovenian Research Agency and research Grants
No.~J1-9124
and
No.~P1-0135;
Agencia Estatal de Investigacion, Spain
Grant No.~RYC2020-029875-I
and
Generalitat Valenciana, Spain
Grant No.~CIDEGENT/2018/020
Ministry of Science and Technology and research Grants
No.~MOST106-2112-M-002-005-MY3
and
No.~MOST107-2119-M-002-035-MY3,
and the Ministry of Education (Taiwan);
Thailand Center of Excellence in Physics;
TUBITAK ULAKBIM (Turkey);
National Research Foundation of Ukraine, project No.~2020.02/0257,
and
Ministry of Education and Science of Ukraine;
the U.S. National Science Foundation and research Grants
No.~PHY-1913789 
and
No.~PHY-2111604, 
and the U.S. Department of Energy and research Awards
No.~DE-AC06-76RLO1830, 
No.~DE-SC0007983, 
No.~DE-SC0009824, 
No.~DE-SC0009973, 
No.~DE-SC0010007, 
No.~DE-SC0010073, 
No.~DE-SC0010118, 
No.~DE-SC0010504, 
No.~DE-SC0011784, 
No.~DE-SC0012704, 
No.~DE-SC0019230, 
No.~DE-SC0021274, 
No.~DE-SC0022350, 
No.~DE-SC0023470; 
and
the Vietnam Academy of Science and Technology (VAST) under Grant No.~DL0000.05/21-23.

These acknowledgements are not to be interpreted as an endorsement of any statement made
by any of our institutes, funding agencies, governments, or their representatives.

We thank the SuperKEKB team for delivering high-luminosity collisions;
the KEK cryogenics group for the efficient operation of the detector solenoid magnet;
the KEK computer group and the NII for on-site computing support and SINET6 network support;
and the raw-data centers at BNL, DESY, GridKa, IN2P3, INFN, and the University of Victoria for offsite computing support.

\clearpage
\end{document}